\documentclass[twocolumn,aps,superscriptaddress,showpacs,floatfix]{revtex4}
\usepackage{amssymb}
\usepackage{amsmath}
\usepackage{graphicx}
\usepackage[normalem]{ulem}
\usepackage[dvips]{color}
\usepackage{bm}
\usepackage{longtable}
\usepackage{array}
\usepackage{etoolbox}
\usepackage{setspace}
\usepackage[colorlinks,linkcolor=blue,anchorcolor=blue,citecolor=blue,urlcolor=blue]{hyperref}

\renewcommand\sout{\bgroup \color{red} \ULdepth=-.5ex \ULset}

\begin{document}

\title{Systematic study of $\alpha$ decay for isomer related nuclei within a two-potential approach}

\author{Xiao-Dong Sun}
\affiliation{School of Math and Physics, University of South China, 421001 Hengyang, People's Republic of China}
\author{Ping Guo}
\affiliation{School of Math and Physics, University of South China, 421001 Hengyang, People's Republic of China}
\author{Xiao-Hua Li\footnote{%
Corresponding author:lixiaohuaphysics@126.com }}
\affiliation{School of Nuclear Science and Technology, University of South China, 421001 Hengyang, People's Republic of China}
\affiliation{Cooperative Innovation Center for Nuclear Fuel Cycle Technology $\&$ Equipment, University of South China, 421001 Hengyang, People's Republic of China}


\begin{abstract}

$\alpha$ decay occurs in both ground states and isomers of nuclei. In this work, we use the two-potential approach to systematically study whether isomeric states play a key role on $\alpha$ particle clustering or not. The results indicate the ratios of $\alpha$ decay preformation probabilities of isomers to ground states are found to be around 1. 

\end{abstract}

\maketitle

\section{Introduction}

Since the first observation of nuclear isomer $^{234m}_{91}Pa$ in 1921, many basic knowledge about isomeric states have being obtained~\cite{Gol51,Bai53,Cam11,Apr05,Blo08,Zha12,Cwi05,And99,Pol99}. The nuclear decay modes depend on its energy level, spin and parity. Light isomers usually go to stability through $\gamma$ decay or $\beta$ decay, while some heavier ones also tend to $\alpha$ decay or even spontaneous fission~\cite{Aud12}. Some $\alpha$ decay chains of isomeric states occur in heavy nuclei~\cite{Hof79,Hof00}. The nuclear isomers in super-heavy elements can be treated as steeping stones towards the island of stability~\cite{Her06}. However it is doubtful that whether the single nucleon excitation or multiquasiparticle states play a key role on nuclear clustering behavior. 

$\alpha$ decay has long been a open source providing nuclear structural information, for example the nuclear shell effect, properties of ground state, energy levels and low lying states, nuclear shape coexistence and so on~\cite{Wau94,And00}. For nuclei far away from the beta-stability line with very short life-time, $\alpha$ decay is one of the most powerful research tools, especially in the super-heavy element~\cite{San09,Sam07}. In addition, experimental researches on the natural occurring long-life $\alpha$ decay nucleus~\cite{De03} and the clock in the solar system $^{146}$Sm are one of hot topics in the field of nuclear physics~\cite{Kin12}. 

Recently, the calculated $\alpha$ decay half-lives using the generalized liquid drop model, the density-dependent cluster model, the unified model for $\alpha$ decay and $\alpha$ capture and so on, can well reproduce the experimental data~\cite{Guo15,Xu05,Qian14,Den15,Buck91,Gon93,qi14}. We have also preformed a systematic calculation of $\alpha$ decay half-lives of even-even nuclei (with the even number of protons and neutrons) using the two-potential approach considering the variations of the $\alpha$ particle preformation probabilities due to the nuclear shell closures~\cite{Sun15}. In this work, we expand this model into $\alpha$ decay of odd-A, odd-odd nuclei from both the isomeric states and ground states. In theory, we assume that the $\alpha$ particle preforms in the decaying nucleus and the orders of the $\alpha$ particle preformation probabilities are related with nuclear structure. The effective $\alpha$ particle preformation probabilities can be extracted from the ratio of calculated $\alpha$ decay half-lives to experimental data. We extract the $\alpha$ particle preformation probabilities ratios of the isomers to the corresponding ground states to study the effect of isomeric states on the $\alpha$ particle preformation probability in the region of heavy and super-heavy elements. 

This article is organized as follows. In Section II, the theoretical framework for the calculation of the $\alpha$ decay half-lives is briefly described. In Section III, the systematic comparations of the $\alpha$ particle preformation probabilities between the isomeric and ground states are preformed, and the preformation probabilities of nuclear isomers are discussed in detail. A brief summary is given in Section IV.

\section{Theoretical framework}

The two-potential approach for metastable states has been widely used to calculate $\alpha$ decay half-lives $T_{1/2}$, which determined by the decay constant $\lambda$. It can be written as
\begin{eqnarray}\label{1}
T_{1/2}=\frac{ln2}{\lambda}.
\end{eqnarray}
The decay constant $\lambda$ depending on the $\alpha$ particle preformation probability $P_\alpha$, the penetration probability $P$, the normalized factor $F$, can be expressed as
\begin{eqnarray}\label{2}
\lambda=P_{\alpha}FP.
\end{eqnarray}
Under this model the $\alpha$ particle assault frequency is obtained by normalizing the outgoing wave functions, while the generalized liquid drop model take it by a classical apporaximation~\cite{Guo15} and the effective liquid drop model put it as a adjustable parameter~\cite{Gon93}. The normalized factor F, expressing the assult frequency, can be approximated as
\begin{eqnarray}\label{3}
F\int_{r_1}^{r_2}\frac{\mathit{d}r}{2k(r)}=1,
\end{eqnarray}
where $r$ is the mass center distance between the preformed $\alpha$ particle and the daughter nucleus. The $r_1$, $r_2$ and following $r_3$ are the classical turning points. The classical turning points satisfy the condition $V(r_1)=V(r_2)=V(r_3)=Q_\alpha$. $k(r)=\sqrt{\frac{2\mu}{\hbar^2}\mid Q_\alpha-V(r)\mid}$ is the wave number. $\mu$ is the reduced mass of the $\alpha$ particle and daughter nucleus. $V(r)$ and $Q_\alpha$ are the height of $\alpha$-core potential and $\alpha$ decay energy, respectively. The penetration probability $P$, which is calculated by the WKB approximation~\cite{Gam28,Gur28}, can be expressed as
\begin{eqnarray}\label{4}
P=exp[-\frac{2}{\hbar}\int_{r_2}^{r_3}k(r)\mathit{d}r].
\end{eqnarray}
The preformed $\alpha$ particle is tightly bound to the daughter nucleus in the inner region $(r_1<r<r_2)$ by the strong interaction, except it penetrates the barrier in the region $(r_2<r<r_3)$ with a very small probability and emitted from the decaying nucleus. 

The potential between the preformed $\alpha$ particle and the daughter nucleus, including nuclear, Coulomb and centrifugal potential barrier, can be written as
\begin{eqnarray}\label{5}
V(r)=V_N(r)+V_C(r)+V_l(r),
\end{eqnarray}
where $V_N(r)$ represents nuclear potential. In this work, we choose a type of $cosh$ for the nuclear potential~\cite{Buc90}. It can be expressed as
\begin{eqnarray}\label{6}
V_N(r)=-V_0\frac{1+cosh(R/a)}{cosh(r/a)+cosh(R/a)},
\end{eqnarray}
where $V_0$ and $a$ are the depth and diffuseness of the nuclear potential, respectively. $V_C(r)$ is the Coulomb potential and is taken as the potential of a uniformly charged sphere with sharp radius $R$, which can be expressed as
\begin{eqnarray}\label{7}
V_C(r)= \left \{
\begin{aligned}
\frac{Z_dZ_{\alpha}\mathit{e}^2}{2R}[3-(\frac{r}{R})^2]~~~r<R\\
\frac{Z_dZ_{\alpha}\mathit{e}^2}{2r}~~~~~~~~~~~~~~~~r>R,
\end{aligned}
\right.
\end{eqnarray}
where $Z_d$ and $Z_{\alpha}$ are proton number of the daughter nucleus and the $\alpha$ particle, respectively. The sharp radius $R$ is given by
\begin{eqnarray}\label{8}
R=1.28A^{1/3}-0.76+0.8A^{-1/3}.
\end{eqnarray}
This empirical formula is commonly used to calculate $\alpha$ decay half-lives~\cite{Roy00}, which is derived from the nuclear droplet model and the proximity energy. The last part, centrifugal potential can be estimated by
\begin{eqnarray}\label{9}
V_l(r)=\frac{l(l+1) \hbar^2}{2\mu r^2},
\end{eqnarray}
where $l$ is the orbital angular momentum taken away by $\alpha$ particle. Most of even-even nuclei, part of odd-mass and odd-odd nuclei occurs the favored $\alpha$ decay with $l=0$. While the unfavored decays with $l\neq 0$ are hindered by the additional centrifugal potential. 

In our previous paper we have fitted the experimental $\alpha$ decay half-lives of 164 even-even nuclei obtaining a set of parameters, which is $a=0.5654~fm, V_0=189.53~MeV, P_0=0.7$~\cite{Sun15}. The diffuseness $a$ and depth $V_0$ of the nuclear potential within this work remain unchanged. The $\alpha$ particle preformation probability $P_\alpha$ will systematically varies due to the nuclear shell effect and varies smoothly in the open-shell region~\cite{Qia13}. The researches also depict that a smaller $\alpha$ particle preformation probability is required for odd-A nuclei then even-even nuclei. So we choose the $\alpha$ particle preformation factor $P_0=0.7$ for even-even nuclei, $P_0=0.24$ for odd-A nuclei, $P_0=0.08$ for odd-odd nuclei by fitting the experimental half-lives. 

\section{Results and discussions}

We have calculated the $\alpha$ decay half-lives of the nuclear isomers and the corresponding ground states, and the $\alpha$ particle preformation probabilities ratio of nuclear isomers to corresponding ground states. The detailed numerical results are presented in Table \ref{Tab1}, where the experimental data are taken from Ref.~\cite{Aud12}. The first and second columns denote the $\alpha$ transition and the spin-parity transition, respectively. Elements with upper suffix $"m", "n", "p", "q"$ indicate assignments to excited isomeric states(with half-lives greater than 100 ns). The $"()"$ in spin or parity means uncertain, and the values with $"\#"$ are estimated from the trends in neighboring nuclides with the same Z and N parities. The next two columns stand for the number of protons and neutrons of the parent nucleus, respectively. The fifth column $l_{min}$ denotes the minimal angular momentum quantum number carried out by the emitted $\alpha$ particle obeying the spin-parity rule. The sixth column represents the decay energy $Q_\alpha$ in unit of MeV. The next two columns show the experimental half-lives and calculated results in unit of second, respectively. The last column $P^*_\alpha/P_\alpha$ denotes the $\alpha$ particle preformation probability ratios. The superscript $*$ refer to the nuclear isomers. 

The effective $\alpha$ particle preformation probability is defined as $P_\alpha=T^{cal}_{1/2}/T^{exp}_{1/2}$. The superscript $exp$ and $cal$ represent experimental data and calculated values, respectively. If the authentic $\alpha$ particle preformation probability is less than the hypothetical preformation factor $P_0$, the calculated $\alpha$ decay half-life $T^{cal}_{1/2}$ could be shorter than the experimental data $T^{exp}_{1/2}$ and the effective preformation probability $P_\alpha$ could be small. This case usually takes place near the nuclear shell closures. In conclusion, the authentic $\alpha$ particle preformation probability equals to the product of preformation factor $P_0$ and effective preformation probability $P_\alpha$. It is worth mention that the $\alpha$ transitions in odd lines only include the ground states in Table \ref{Tab1}, while the others in even lines have to contain the nuclear isomers. So the preformation probability ratio $P^*_\alpha/P_\alpha$ imply the influence of isomeric states on the $\alpha$ particle preformation probability. The ratio is greater than 1 meaning the preformation probability increases due to the isomer, and vice versa. 

\begingroup
\renewcommand*{\arraystretch}{1.3}
\begin{longtable*}{ccccccccc}
\caption{Calculations of $\alpha$ decay half-lives and the $\alpha$ particle preformation probabilities ratio of nuclear isomer to ground state.}
\label{Tab1} \\
\hline 
$\alpha$ transition & $I^{\pi}_i\to I^{\pi}_j$ & $Z_p$ & $N_p$ & $l_{min}$ & $Q_\alpha$ & $T^{exp}_{1/2}(s)$ & $T^{cal}_{1/2}(s)$ & $P^*_\alpha/P_\alpha$ \\ 
\hline
\endfirsthead
\multicolumn{9}{c}%
{{\tablename\ \thetable{} -- continued from previous page}} \\
\hline 
$\alpha$ transition & $I^{\pi}_i\to I^{\pi}_j$ & $Z_p$ & $N_p$ & $l_{min}$ & $Q_\alpha$ & $T^{exp}_{1/2}(s)$ & $T^{cal}_{1/2}(s)$ & $P^*_\alpha/P_\alpha$ \\ 
\hline
\endhead
\hline \multicolumn{9}{r}{{Continued on next page}} \\
\endfoot
\hline \hline
\endlastfoot
$^{149}$Tb$\to^{145}$Eu&1/2$^+$$\to$5/2$^+$&65&84&2&4.078&$8.88\times10^{4}$&$7.38\times10^{4}$&-\\
$^{149m}$Tb$\to^{145}$Eu&11/2$^-$$\to$5/2$^+$ &65&84&3&4.114&$1.13\times10^{6}$&$9.62\times10^{4}$&0.102\\
$^{153}$Tm$\to^{149}$Ho&(11/2$^-$)$\to$(11/2$^-$)&69&84&0&5.248&$1.63\times10^{0}$&$3.40\times10^{0}$&-\\
$^{153m}$Tm$\to^{149m}$Ho&(1/2$^+$)$\to$(1/2$^+$) &69&84&0&5.242&$2.72\times10^{0}$&$3.36\times10^{0}$&0.592\\
$^{155}$Lu$\to^{151}$Tm&(11/2$^-$)$\to$(11/2$^-$)&71&84&0&5.803&$7.62\times10^{-2}$&$2.24\times10^{-1}$&-\\
$^{155m}$Lu$\to^{151m}$Tm&(1/2$^+$)$\to$(1/2$^+$) &71&84&0&5.730&$1.82\times10^{-1}$&$2.67\times10^{-1}$&0.5\\
$^{155n}$Lu$\to^{151}$Tm&25/2$^{-\#}$$\to$(11/2$^-$)&71&84&8&7.584&$2.69\times10^{-3}$&$4.03\times10^{-4}$&0.051\\
$^{156}$Hf$\to^{152}$Yb&0$^+$$\to$0$^+$&72&84&0&6.025&$2.37\times10^{-2}$&$2.03\times10^{-2}$&-\\
$^{156m}$Hf$\to^{152}$Yb&8$^+$$\to$0$^+$&72&84&8&7.985&$4.80\times10^{-4}$&$2.57\times10^{-5}$&0.062\\
$^{158}$W$\to^{154}$Hf&0$^+$$\to$0$^+$&74&84&0&6.605&$1.25\times10^{-3}$&$1.13\times10^{-3}$&-\\
$^{158m}$W$\to^{154}$Hf&8$^+$$\to$0$^+$&74&84&8&8.495&$1.43\times10^{-4}$&$8.42\times10^{-6}$&0.065\\

$^{152}$Ho$\to^{148}$Tb&2$^-$$\to$2$^-$&67&85&0&4.507&$1.35\times10^{3}$&$4.16\times10^{3}$&-\\
$^{152m}$Ho$\to^{148m}$Tb&9$^+$$\to$(9)$^+$&67&85&0&4.577&$4.63\times10^{2}$&$1.76\times10^{3}$&1.235\\
$^{154}$Tm$\to^{150}$Ho&(2$^-$)$\to$2$^-$&69&85&0&5.094&$1.49\times10^{1}$&$5.62\times10^{1}$&-\\
$^{154m}$Tm$\to^{150m}$Ho&(9$^+$)$\to$(9)$^+$&69&85&0&5.175&$5.64\times10^{0}$&$1.94\times10^{1}$&0.91\\
$^{158}$Ta$\to^{154}$Lu&(2$^-$)$\to$(2$^-$)&73&85&0&6.125&$5.10\times10^{-2}$&$5.33\times10^{-1}$&-\\
$^{158m}$Ta$\to^{154m}$Lu&(9$^+$)$\to$(9$^+$)&73&85&0&6.205&$3.79\times10^{-2}$&$9.34\times10^{-2}$&0.236\\

$^{162}$Re$\to^{158}$Ta&(2$^-$)$\to$(2$^-$)&75&87&0&6.245&$1.14\times10^{-1}$&$4.51\times10^{-1}$&-\\
$^{162m}$Re$\to^{158m}$Ta&(9$^+$)$\to$(9$^+$)&75&87&0&6.275&$8.46\times10^{-2}$&$3.90\times10^{-1}$&1.165\\
$^{163}$Re$\to^{159}$Ta&1/2$^+$$\to$1/2$^+$&75&88&0&6.012&$1.22\times10^{0}$&$1.19\times10^{0}$&-\\
$^{163m}$Re$\to^{159m}$Ta&11/2$^-$$\to$11/2$^-$&75&88&0&6.068&$3.24\times10^{-1}$&$7.14\times10^{-1}$&2.252\\

$^{166}$Ir$\to^{162}$Re&(2$^-$)$\to$(2$^-$)&77&89&0&6.725&$1.13\times10^{-2}$&$6.59\times10^{-2}$&-\\
$^{166m}$Ir$\to^{162m}$Re&(9$^+$)$\to$(9$^+$)&77&89&0&6.725&$1.54\times10^{-2}$&$6.46\times10^{-2}$&0.719\\
$^{167}$Ir$\to^{163}$Re&1/2$^+$$\to$1/2$^+$&77&90&0&6.504&$6.81\times10^{-2}$&$1.03\times10^{-1}$&-\\
$^{167m}$Ir$\to^{163m}$Re&11/2$^-$$\to$11/2$^-$&77&90&0&6.561&$2.86\times10^{-2}$&$6.64\times10^{-2}$&1.542\\

$^{168}$Ir$\to^{164}$Re&(2$^-$)$\to$(2$^-$)&77&91&0&6.375&$2.30\times10^{-1}$&$1.10\times10^{0}$&-\\
$^{168m}$Ir$\to^{164m}$Re&(9$^+$)$\to$(9$^+$)&77&91&0&6.485&$2.12\times10^{-1}$&$3.54\times10^{-1}$&0.348\\
$^{170}$Au$\to^{166}$Ir&(2$^-$)$\to$(2$^-$)&79&91&0&7.175&$2.64\times10^{-3}$&$1.01\times10^{-2}$&-\\
$^{170m}$Au$\to^{166m}$Ir&(9$^+$)$\to$(9$^+$)&79&91&0&7.285&$1.48\times10^{-3}$&$5.96\times10^{-3}$&1.049\\
$^{169}$Ir$\to^{165}$Re&(1/2$^+$)$\to$(1/2$^+$)&77&92&0&6.141&$7.84\times10^{-1}$&$2.28\times10^{0}$&-\\
$^{169m}$Ir$\to^{165m}$Re&(11/2$^-$)$\to$(11/2$^-$)&77&92&0&6.266&$3.90\times10^{-1}$&$1.02\times10^{0}$&0.894\\

$^{173}$Au$\to^{169}$Ir&(1/2$^+$)$\to$(1/2$^+$)&79&94&0&6.837&$2.91\times10^{-2}$&$4.29\times10^{-2}$&-\\
$^{173m}$Au$\to^{169m}$Ir&(11/2$^-$)$\to$(11/2$^-$) &79&94&0&6.896&$1.57\times10^{-2}$&$4.52\times10^{-2}$&1.946\\

$^{172}$Ir$\to^{168}$Re&(3$^+$)$\to$(7$^+$)&77&95&4&5.985&$2.20\times10^{2}$&$2.25\times10^{2}$&-\\
$^{172m}$Ir$\to^{168}$Re&(7$^+$)$\to$(7$^+$)&77&95&0&6.125&$8.70\times10^{0}$&$7.39\times10^{0}$&0.83\\
$^{175}$Au$\to^{171}$Ir&1/2$^+$$\to$1/2$^+$&79&96&0&6.575&$2.16\times10^{-1}$&$3.27\times10^{-1}$&-\\
$^{175m}$Au$\to^{171m}$Ir&(11/2$^-$)$\to$(11/2$^-$)&79&96&0&6.585&$1.79\times10^{-1}$&$3.33\times10^{-1}$&1.232\\
$^{177}$Tl$\to^{173}$Au&(1/2$^+$)$\to$(1/2$^+$)&81&96&0&7.066&$2.47\times10^{-2}$&$5.21\times10^{-2}$&-\\
$^{177m}$Tl$\to^{173m}$Au&(11/2$^-$)$\to$(11/2$^-$)&81&96&0&7.654&$3.67\times10^{-4}$&$5.51\times10^{-4}$&0.709\\

$^{174}$Ir$\to^{170}$Re&(3$^+$)$\to$(5$^+$)&77&97&2&5.624&$1.58\times10^{3}$&$2.04\times10^{3}$&-\\
$^{174m}$Ir$\to^{170}$Re&(7$^+$)$\to$(5$^+$)&77&97&2&5.817&$1.96\times10^{2}$&$2.43\times10^{2}$&0.962\\
$^{177}$Au$\to^{173}$Ir&(1/2$^+$,3/2$^+$)$\to$(1/2$^+$,3/2$^+$)&79&98&0&6.298&$3.65\times10^{0}$&$3.26\times10^{0}$&-\\
$^{177m}$Au$\to^{173m}$Ir&11/2$^-$$\to$(11/2$^-$)&79&98&0&6.261&$1.79\times10^{0}$&$5.47\times10^{0}$&3.423\\

$^{185}$Pb$\to^{181}$Hg&3/2$^-$$\to$1/2$^{(-\#)}$&82&103&2&6.695&$1.85\times10^{1}$&$3.22\times10^{0}$&-\\
$^{185m}$Pb$\to^{181m}$Hg&13/2$^+$$\to$13/2$^+$&82&103&0&6.555&$8.14\times10^{0}$&$5.69\times10^{0}$&4.022\\

$^{185}$Hg$\to^{181}$Pt&1/2$^-$$\to$1/2$^-$&80&105&0&5.773&$8.18\times10^{2}$&$1.31\times10^{3}$&-\\
$^{185m}$Hg$\to^{181m}$Pt&13/2$^+$$\to$(7/2)$^-$&80&105&3&5.760&$7.20\times10^{4}$&$5.02\times10^{3}$&0.044\\
$^{187}$Pb$\to^{183}$Hg&3/2$^-$$\to$1/2$^-$&82&105&2&6.393&$1.60\times10^{2}$&$3.98\times10^{1}$&-\\
$^{187m}$Pb$\to^{183m}$Hg&13/2$^+$$\to$13/2$^{+\#}$&82&105&0&6.208&$1.53\times10^{2}$&$1.64\times10^{2}$&4.32\\
$^{187}$Tl$\to^{183}$Au&(1/2$^+$)$\to$(1/2)$^+$&81&106&0&5.248&$1.70\times10^{5}$&$1.73\times10^{6}$&-\\
$^{187m}$Tl$\to^{183m}$Au&(9/2$^-$)$\to$(11/2)$^-$&81&106&2&5.425&$1.56\times10^{4}$&$3.30\times10^{5}$&2.082\\

$^{187}$Hg$\to^{183}$Pt&3/2$^{(-)}$$\to$1/2$^-$&80&107&2&5.229&$<9.50\times10^{7}$&$1.26\times10^{6}$&-\\
$^{187m}$Hg$\to^{183m}$Pt&13/2$^+$$\to$(7/2)$^-$&80&107&3&5.254&$<5.76\times10^{7}$&$1.44\times10^{6}$&1.88\\
$^{189}$Pb$\to^{185}$Hg&3/2$^-$$\to$1/2$^-$&82&107&2&5.871&$1.26\times10^{4}$&$8.43\times10^{3}$&-\\
$^{189m}$Pb$\to^{185m}$Hg&13/2$^+$$\to$13/2$^+$&82&107&0&5.807&$>3.90\times10^{3}$&$9.27\times10^{3}$&3.557\\
$^{190}$Bi$\to^{186}$Tl&(3$^+$)$\to$(2$^-$)&83&107&1&6.863&$8.18\times10^{0}$&$3.35\times10^{0}$&-\\
$^{190m}$Bi$\to^{186m}$Tl&(10$^-$)$\to$(7$^+$)&83&107&3&6.965&$8.86\times10^{0}$&$3.68\times10^{0}$&1.016\\
$^{192}$At$\to^{188}$Bi&3$^{+\#}$$\to$3$^{+\#}$&85&107&0&7.700&$1.15\times10^{-2}$&$2.79\times10^{-2}$&-\\
$^{192m}$At$\to^{188m}$Bi&(9$^-$,10$^-$)$\to$(10$^-$)&85&107&0&7.545&$8.80\times10^{-2}$&$7.75\times10^{-2}$&0.364\\

$^{191}$Pb$\to^{187}$Hg&(3/2$^-$)$\to$3/2$^{(-)}$&82&109&0&5.453&$1.56\times10^{4}$&$3.64\times10^{5}$&-\\
$^{191m}$Pb$\to^{187m}$Hg&13/2$^{(+)}$$\to$13/2$^+$&82&109&0&5.404&$6.54\times10^{5}$&$6.67\times10^{5}$&0.044\\
$^{192}$Bi$\to^{188}$Tl&(3$^+$)$\to$(2$^-$)&83&109&1&6.371&$2.77\times10^{2}$&$2.51\times10^{2}$&-\\
$^{192m}$Bi$\to^{188m}$Tl&(10$^-$)$\to$(9$^-$)&83&109&2&6.513&$3.84\times10^{2}$&$9.31\times10^{1}$&0.267\\
$^{194}$At$\to^{190}$Bi&(4$^-$,5$^-$)$\to$(3$^+$)&85&109&1&7.462&$2.53\times10^{-1}$&$1.69\times10^{-1}$&-\\
$^{194m}$At$\to^{190m}$Bi&(9$^-$,10$^-$)$\to$(10$^-$)&85&109&0&7.335&$3.10\times10^{-1}$&$3.57\times10^{-1}$&1.723\\
$^{195}$Rn$\to^{191}$Po&(3/2$^-$)$\to$(3/2$^-$)&86&109&0&7.694&$7.00\times10^{-3}$&$1.94\times10^{-2}$&-\\
$^{195m}$Rn$\to^{191m}$Po&(13/2$^+$)$\to$(13/2$^+$)&86&109&0&7.714&$6.00\times10^{-3}$&$1.90\times10^{-2}$&1.139\\

$^{191}$Hg$\to^{187}$Pt&3/2$^{(-)}$$\to$3/2$^-$&80&111&0&3.667&$>5.88\times10^{10}$&$4.70\times10^{16}$&-\\
$^{191m}$Hg$\to^{187m}$Pt&13/2$^{(+)}$$\to$(11/2$^+$)&80&111&2&3.616&$>6.10\times10^{10}$&$2.80\times10^{17}$&5.757\\
$^{194}$Bi$\to^{190}$Tl&(3$^+$)$\to$2$^{(-)}$&83&111&1&5.915&$2.07\times10^{4}$&$2.06\times10^{4}$&-\\
$^{194n}$Bi$\to^{190m}$Tl&(10$^-$)$\to$9$^-$&83&111&3&6.015&$5.75\times10^{4}$&$2.17\times10^{4}$&0.379\\
$^{195}$Po$\to^{191}$Pb&(3/2$^-$)$\to$(3/2$^-$)&84&111&0&6.755&$4.94\times10^{0}$&$4.92\times10^{0}$&-\\
$^{195m}$Po$\to^{191m}$Pb&(13/2$^+$)$\to$13/2$^{(+)}$&84&111&0&6.840&$2.13\times10^{0}$&$2.28\times10^{0}$&1.07\\
$^{197}$Rn$\to^{193}$Po&(3/2$^-$)$\to$(3/2$^-$)&86&111&0&7.415&$5.40\times10^{-2}$&$1.66\times10^{-1}$&-\\
$^{197m}$Rn$\to^{193m}$Po&(13/2$^+$)$\to$(13/2$^+$)&86&111&0&7.505&$2.56\times10^{-2}$&$7.94\times10^{-2}$&1.007\\

$^{197}$At$\to^{193}$Bi&(9/2$^-$)$\to$(9/2$^-$)&85&112&0&7.108&$4.04\times10^{-1}$&$5.95\times10^{-1}$&-\\
$^{197m}$At$\to^{193m}$Bi&(1/2$^+$)$\to$(1/2$^+$)&85&112&0&6.846&$2.00\times10^{0}$&$5.70\times10^{0}$&1.936\\
$^{198}$At$\to^{194}$Bi&(3$^+$)$\to$(3$^+$)&85&113&0&6.895&$<4.48\times10^{0}$&$1.09\times10^{1}$&-\\
$^{198m}$At$\to^{194m}$Bi&(10$^-$)$\to$(10$^-$)&85&113&0&6.995&$<1.27\times10^{0}$&$6.46\times10^{0}$&2.1\\
$^{197}$Po$\to^{193}$Pb&(3/2$^-$)$\to$(3/2$^-$)&84&113&0&6.405&$1.22\times10^{2}$&$1.03\times10^{2}$&-\\
$^{197m}$Po$\to^{193m}$Pb&(13/2$^+$)$\to$13/2$^+$&84&113&0&6.505&$3.07\times10^{1}$&$4.51\times10^{1}$&1.732\\
$^{200}$Fr$\to^{196}$At&(3$^+$)$\to$(3$^+$)&87&113&0&7.615&$4.90\times10^{-2}$&$3.09\times10^{-1}$&-\\
$^{200m}$Fr$\to^{196m}$At&10$^{-\#}$$\to$(10$^-$)&87&113&0&7.705&$1.90\times10^{-1}$&$1.15\times10^{-1}$&0.096\\

$^{201}$Fr$\to^{197}$At&(9/2$^-$)$\to$(9/2$^-$)&87&114&0&7.515&$6.20\times10^{-2}$&$1.49\times10^{-1}$&-\\
$^{201m}$Fr$\to^{197m}$At&(1/2$^+$)$\to$(1/2$^+$)&87&114&0&7.608&$2.70\times10^{-2}$&$7.18\times10^{-2}$&1.109\\
$^{199}$Po$\to^{195}$Pb&3/2$^{-\#}$$\to$3/2$^{-\#}$&84&115&0&6.074&$4.38\times10^{3}$&$3.35\times10^{3}$&-\\
$^{199m}$Po$\to^{195m}$Pb&13/2$^{(+)}$$\to$13/2$^{(+)}$&84&115&0&6.181&$1.04\times10^{3}$&$9.84\times10^{2}$&1.233\\
$^{200}$At$\to^{196}$Bi&(3$^+$)$\to$(3$^+$)&85&115&0&6.596&$8.31\times10^{1}$&$1.37\times10^{2}$&-\\
$^{200m}$At$\to^{196m}$Bi&(7$^+$)$\to$(7$^+$)&85&115&0&6.543&$1.09\times10^{2}$&$2.21\times10^{2}$&1.221\\
$^{200n}$At$\to^{196n}$Bi&(10$^-$)$\to$(10$^-$)&85&115&0&6.669&$7.62\times10^{1}$&$8.72\times10^{1}$&0.693\\
$^{203}$Ra$\to^{199}$Rn&(3/2$^-$)$\to$(3/2$^-$)&88&115&0&7.745&$3.60\times10^{-2}$&$6.54\times10^{-2}$&-\\
$^{203m}$Ra$\to^{199m}$Rn&(13/2$^+$)$\to$(13/2$^+$)&88&115&0&7.765&$2.50\times10^{-2}$&$6.44\times10^{-2}$&1.417\\

$^{201}$Po$\to^{197}$Pb&3/2$^-$$\to$3/2$^-$&84&117&0&5.799&$8.28\times10^{4}$&$6.27\times10^{4}$&-\\
$^{201m}$Po$\to^{197m}$Pb&13/2$^+$$\to$13/2$^+$&84&117&0&5.903&$2.24\times10^{4}$&$1.53\times10^{4}$&0.901\\
$^{202}$At$\to^{198}$Bi&(2$^+$,3$^+$)$\to$(2$^+$ 3$^+$)&85&117&0&6.353&$4.97\times10^{2}$&$1.46\times10^{3}$&-\\
$^{202m}$At$\to^{198m}$Bi&(7$^+$)$\to$(7$^+$)&85&117&0&6.259&$2.09\times10^{3}$&$4.00\times10^{3}$&0.649\\
$^{202n}$At$\to^{198n}$Bi&(10$^-$)$\to$(10$^-$)&85&117&0&6.402&$4.79\times10^{2}$&$7.81\times10^{2}$&0.853\\
$^{203}$Rn$\to^{199}$Po&3/2$^{-\#}$$\to$3/2$^{-\#}$&86&117&0&6.630&$6.67\times10^{1}$&$8.35\times10^{1}$&-\\
$^{203m}$Rn$\to^{199m}$Po&13/2$^{(+)}$$\to$13/2$^{(+)}$&86&117&0&6.680&$3.59\times10^{1}$&$5.24\times10^{1}$&1.166\\
$^{204}$Fr$\to^{200}$At&(3$^+$)$\to$(3$^+$)&87&117&0&7.170&$1.82\times10^{0}$&$6.68\times10^{0}$&-\\
$^{204m}$Fr$\to^{200m}$At&(7$^+$)$\to$(7$^+$)&87&117&0&7.108&$2.56\times10^{0}$&$1.28\times10^{1}$&1.364\\
$^{204n}$Fr$\to^{200n}$At&(10$^-$)$\to$(10$^-$)&87&117&0&7.154&$1.08\times10^{0}$&$6.63\times10^{0}$&1.228\\
$^{206}$Ac$\to^{202}$Fr&(3$^+$)$\to$(3$^+$)&89&117&0&7.945&$2.50\times10^{-2}$&$1.59\times10^{-1}$&-\\
$^{206m}$Ac$\to^{202m}$Fr&(10$^-$)$\to$(10$^-$)&89&117&0&7.904&$4.10\times10^{-2}$&$1.36\times10^{-1}$&0.523\\

$^{203}$Po$\to^{199}$Pb&5/2$^-$$\to$3/2$^-$&84&119&2&5.496&$2.00\times10^{6}$&$2.70\times10^{6}$&-\\
$^{203m}$Po$\to^{199m}$Pb&13/2$^+$$\to$(13/2$^+$)&84&119&0&5.709&$1.13\times10^{5}$&$1.33\times10^{5}$&0.877\\
$^{206}$Fr$\to^{202}$At&(2$^+$,3$^+$)$\to$(2$^+$ 3$^+$)&87&119&0&6.924&$3.81\times10^{1}$&$4.40\times10^{1}$&-\\
$^{206m}$Fr$\to^{202m}$At&(7$^+$)$\to$(7$^+$)&87&119&0&6.928&$3.81\times10^{1}$&$5.25\times10^{1}$&1.193\\
$^{206n}$Fr$\to^{202n}$At&(10$^-$)$\to$(10$^-$)&87&119&0&7.068&$1.40\times10^{1}$&$1.54\times10^{1}$&0.954\\

$^{209}$Ra$\to^{205}$Rn&5/2$^-$$\to$5/2$^-$&88&121&0&7.135&$5.23\times10^{0}$&$7.54\times10^{0}$&-\\
$^{209m}$Ra$\to^{205}$Rn&13/2$^+$$\to$5/2$^-$&88&121&5&8.015&$1.30\times10^{-4}$&$1.38\times10^{-1}$&736.4\\
$^{213}$Ra$\to^{209}$Rn&1/2$^-$$\to$5/2$^-$&88&125&2&6.862&$2.05\times10^{2}$&$1.37\times10^{2}$&-\\
$^{213m}$Ra$\to^{209m}$Rn&(17/2$^-$)$\to$13/2$^+$&88&125&3&7.456&$3.67\times10^{-1}$&$1.38\times10^{0}$&5.648\\

$^{217}$Pa$\to^{213}$Ac&9/2$^{-\#}$$\to$9/2$^{-\#}$&91&126&0&8.485&$3.48\times10^{-3}$&$2.94\times10^{-3}$&-\\
$^{217m}$Pa$\to^{213}$Ac&29/2$^{+\#}$$\to$9/2$^{-\#}$&91&126&11&10.345&$1.48\times10^{-3}$&$5.18\times10^{-3}$&4.139\\
$^{214}$Ra$\to^{210}$Rn&0$^+$$\to$0$^+$&88&126&0&7.273&$2.46\times10^{0}$&$5.41\times10^{-1}$&-\\
$^{214n}$Ra$\to^{210}$Rn&8$^+$$\to$0$^+$&88&126&8&9.138&$7.48\times10^{-2}$&$8.19\times10^{-4}$&0.0498\\
$^{216}$Th$\to^{212}$Ra&0$^+$$\to$0$^+$&90&126&0&8.073&$2.60\times10^{-2}$&$6.75\times10^{-3}$&-\\
$^{216m}$Th$\to^{212}$Ra&(8$^+$)$\to$0$^+$&90&126&8&10.116&$4.79\times10^{-3}$&$2.03\times10^{-5}$&0.00828\\
$^{218}$U$\to^{214}$Th&0$^+$$\to$0$^+$&92&126&0&8.775&$5.50\times10^{-4}$&$3.44\times10^{-4}$&-\\
$^{218m}$U$\to^{214}$Th&(8$^+$)$\to$0$^+$&92&126&8&10.878&$6.40\times10^{-4}$&$2.58\times10^{-6}$&0.00365\\

$^{210}$Bi$\to^{206}$Tl&1$^-$$\to$0$^-$&83&127&2&5.036&$3.28\times10^{11}$&$5.76\times10^{8}$&-\\
$^{210m}$Bi$\to^{206}$Tl&9$^-$$\to$0$^-$&83&127&10&5.307&$9.59\times10^{13}$&$4.12\times10^{11}$&2.446\\
$^{211}$Po$\to^{207}$Pb&9/2$^+$$\to$1/2$^-$&84&127&5&7.594&$5.16\times10^{-1}$&$4.45\times10^{-2}$&-\\
$^{211m}$Po$\to^{207m}$Pb&(25/2$^+$)$\to$13/2$^+$&84&127&6&7.423&$2.52\times10^{1}$&$5.71\times10^{-1}$&0.263\\
$^{212}$At$\to^{208}$Bi&(1$^-$)$\to$5$^+$&85&127&5&7.816&$3.14\times10^{-1}$&$6.64\times10^{-2}$&-\\
$^{212m}$At$\to^{208}$Bi&9$^{-\#}$$\to$5$^+$&85&127&5&8.039&$1.20\times10^{-1}$&$1.43\times10^{-2}$&0.564\\
$^{214}$Fr$\to^{210}$At&(1$^-$)$\to$(5)$^+$&87&127&5&8.589&$5.00\times10^{-3}$&$2.58\times10^{-3}$&-\\
$^{214m}$Fr$\to^{210}$At&(8$^-$)$\to$(5)$^+$&87&127&3&8.710&$3.35\times10^{-3}$&$1.91\times10^{-4}$&0.11\\
$^{216}$Ac$\to^{212}$Fr&(1$^-$)$\to$5$^+$&89&127&5&9.236&$4.40\times10^{-4}$&$2.13\times10^{-4}$&-\\
$^{216m}$Ac$\to^{212}$Fr&(9$^-$)$\to$5$^+$&89&127&5&9.279&$4.41\times10^{-4}$&$1.97\times10^{-4}$&0.922\\

$^{212}$Po$\to^{208}$Pb&0$^+$$\to$0$^+$&84&128&0&8.954&$2.99\times10^{-7}$&$1.47\times10^{-7}$&-\\
$^{212m}$Po$\to^{208m}$Pb&(18$^+$)$\to$0$^+$&84&128&18&11.877&$4.51\times10^{1}$&$1.72\times10^{-1}$&0.0077\\
$^{217}$Ac$\to^{213}$Fr&9/2$^-$$\to$9/2$^-$&89&128&0&9.832&$6.90\times10^{-8}$&$2.28\times10^{-7}$&-\\
$^{217m}$Ac$\to^{213m}$Fr&(29/2)$^+$$\to$21/2$^-$&89&128&5&10.254&$1.72\times10^{-5}$&$3.52\times10^{-7}$&0.0062\\
$^{212}$Bi$\to^{208}$Tl&1$^{(-)}$$\to$5$^+$&83&129&5&6.207&$1.01\times10^{4}$&$6.34\times10^{3}$&-\\
$^{212m}$Bi$\to^{208}$Tl&(8$^-$,9$^-$)$\to$5$^+$&83&129&3&6.454&$2.24\times10^{3}$&$1.19\times10^{2}$&0.085\\
$^{214}$At$\to^{210}$Bi&1$^-$$\to$1$^-$&85&129&0&8.987&$>5.58\times10^{-7}$&$2.40\times10^{-6}$&-\\
$^{214n}$At$\to^{210n}$Bi&9$^-$$\to$9$^-$&85&129&0&8.949&$7.60\times10^{-7}$&$2.78\times10^{-6}$&0.849\\

$^{216}$At$\to^{212}$Bi&1$^{(-)}$$\to$1$^{(-)}$&85&131&0&7.950&$3.00\times10^{-4}$&$1.83\times10^{-3}$&-\\
$^{216m}$At$\to^{212m}$Bi&9$^{-\#}$$\to$(8$^-$,9$^-$)&85&131&0&7.863&$1.00\times10^{-4}$&$2.41\times10^{-3}$&3.956\\

$^{236}$U$\to^{232}$Th&0$^+$$\to$0$^+$&92&144&0&4.572&$7.39\times10^{14}$&$2.54\times10^{15}$&-\\
$^{236m}$U$\to^{232}$Th&(0$^+$)$\to$0$^+$&92&144&0&7.323&$>1.20\times10^{-6}$&$6.19\times10^{0}$&1499442.686\\
$^{238}$U$\to^{234}$Th&0$^+$$\to$0$^+$&92&146&0&4.270&$1.41\times10^{17}$&$6.80\times10^{17}$&-\\
$^{238m}$U$\to^{234}$Th&0$^+$$\to$0$^+$&92&146&0&6.828&$>5.60\times10^{-5}$&$6.66\times10^{2}$&2464668.069\\
$^{247}$Md$\to^{243}$Es&(7/2$^-$)$\to$(7/2$^+$)&101&146&0&8.765&$1.19\times10^{0}$&$7.73\times10^{-1}$&-\\
$^{247m}$Md$\to^{243}$Es&(1/2$^-$)$\to$(7/2$^+$)&101&146&3&9.025&$3.16\times10^{-1}$&$3.74\times10^{-1}$&1.821\\

$^{247}$Fm$\to^{243}$Cf&(7/2$^+$)$\to$1/2$^{+\#}$&100&147&4&8.255&$<6.20\times10^{1}$&$8.05\times10^{1}$&-\\
$^{247m}$Fm$\to^{243}$Cf&(1/2$^+$)$\to$1/2$^{+\#}$&100&147&0&8.305&$5.10\times10^{0}$&$1.03\times10^{1}$&1.549\\
$^{249}$Md$\to^{245p}$Es&(7/2$^-$)$\to$(7/2$^-$)&101&148&0&8.155&$<3.90\times10^{1}$&$8.60\times10^{1}$&-\\
$^{249m}$Md$\to^{245q}$Es&(1/2$^-$)$\to$(1/2$^-$)&101&148&0&8.205&$1.90\times10^{0}$&$5.20\times10^{1}$&12.406\\
$^{251}$No$\to^{247}$Fm&(7/2$^+$)$\to$(7/2$^+$)&102&149&0&8.755&$9.64\times10^{-1}$&$1.78\times10^{0}$&-\\
$^{251m}$No$\to^{247m}$Fm&(1/2$^+$)$\to$(1/2$^+$)&102&149&0&8.815&$1.02\times10^{0}$&$1.65\times10^{0}$&0.877\\

$^{250}$Fm$\to^{246}$Cf&0$^+$$\to$0$^+$&100&150&0&7.556&$<2.03\times10^{3}$&$1.55\times10^{3}$&-\\
$^{250m}$Fm$\to^{246}$Cf&(8$^-$)$\to$0$^+$&100&150&9&8.755&$>9.60\times10^{0}$&$1.98\times10^{2}$&26.985\\
$^{253}$Lr$\to^{249}$Md&(7/2$^-$)$\to$(7/2$^-$)&103&150&0&8.925&$7.02\times10^{-1}$&$1.43\times10^{0}$&-\\
$^{253m}$Lr$\to^{249m}$Md&(1/2$^-$)$\to$(1/2$^-$)&103&150&0&8.855&$1.47\times10^{0}$&$1.95\times10^{0}$&0.656\\
$^{254}$No$\to^{250}$Fm&0$^+$$\to$0$^+$&102&152&0&8.227&$5.69\times10^{1}$&$2.94\times10^{1}$&-\\
$^{254m}$No$\to^{250}$Fm&(8$^-$)$\to$0$^+$&102&152&9&9.522&$2.65\times10^{3}$&$4.50\times10^{0}$&0.003\\
$^{255m}$Lr$\to^{251}$Md&(7/2$^-$)$\to$(7/2$^-$)&103&152&0&8.594&$2.54\times10^{0}$&$1.16\times10^{1}$&-\\
$^{255n}$Lr$\to^{251}$Md&(25/2$^+$)$\to$(7/2$^-$)&103&152&9&10.018&$>1.09\times10^{0}$&$1.25\times10^{0}$&0.922\\
$^{257}$Db$\to^{253}$Lr&(9/2$^+$)$\to$(7/2$^-$)&105&152&1&9.205&$<2.45\times10^{0}$&$9.75\times10^{-1}$&-\\
$^{257m}$Db$\to^{253m}$Lr&(1/2$^-$)$\to$(1/2$^-$)&105&152&0&9.315&$>7.70\times10^{-1}$&$3.72\times10^{-1}$&1.213\\

$^{254}$Es$\to^{250}$Bk&(7$^+$)$\to$2$^-$&99&155&5&6.615&$2.38\times10^{7}$&$6.55\times10^{8}$&-\\
$^{254m}$Es$\to^{250}$Bk&2$^+$$\to$2$^-$&99&155&1&6.699&$4.42\times10^{7}$&$1.77\times10^{7}$&0.015\\
$^{258}$Md$\to^{254}$Es&8$^{-\#}$$\to$(7$^+$)&101&157&1&7.271&$4.45\times10^{6}$&$3.78\times10^{5}$&-\\
$^{258m}$Md$\to^{254m}$Es&1$^{-\#}$$\to$2$^+$&101&157&1&7.189&$>2.85\times10^{5}$&$9.14\times10^{5}$&37.698\\
$^{265}$Hs$\to^{261}$Sg&3/2$^{+\#}$$\to$(3/2$^+$)&108&157&0&10.470&$1.96\times10^{-3}$&$2.65\times10^{-3}$&-\\
$^{265m}$Hs$\to^{261m}$Sg&9/2$^{+\#}$$\to$(9/2$^+$)&108&157&0&10.416&$3.60\times10^{-4}$&$3.58\times10^{-3}$&7.355\\

$^{271}$Ds$\to^{267}$Hs&13/2$^{-\#}$$\to$5/2$^{+\#}$&110&161&5&10.875&$9.00\times10^{-2}$&$9.39\times10^{-3}$&-\\
$^{271m}$Ds$\to^{267}$Hs&9/2$^{+\#}$$\to$5/2$^{+\#}$&110&161&2&10.945&$1.70\times10^{-3}$&$9.74\times10^{-4}$&5.494\\
$^{273}$Ds$\to^{269}$Hs&13/2$^{-\#}$$\to$9/2$^{+\#}$&110&163&3&11.365&$2.40\times10^{-4}$&$1.59\times10^{-4}$&-\\
$^{273m}$Ds$\to^{269}$Hs&3/2$^{+\#}$$\to$9/2$^{+\#}$&110&163&4&11.565&$1.20\times10^{-1}$&$1.32\times10^{-4}$&0.002\\
\end{longtable*}
\endgroup

Table \ref{Tab1} shows that the values of the preformation probability ratio $P^*_\alpha/P_\alpha$ vary around 1, demonstrating that there are no obvious differences in the $\alpha$ particle preformation probabilities between the nuclear isomers and the corresponding ground states. For most isomers of odd mass nuclei, they have low excited energy of only little hundreds kiloelectron volt above the ground states. This kind of isomer results from the single particle excitation without breaking nucleonic pairs. It is one kind of the spin trap isomers, contiguously occurring below the closed shell, as a example of below the Z=82, and N=126, where the energy levels reduce due to the spin-orbit interaction. There is also the island of isomeric states of odd-odd nuclei in this regions, which is in a similar configuration but more complex due to the proton neutron interaction. The cases of even-even nuclear isomer are rare and interesting, such as the $^{156m}$Hf, $^{158m}$W, $^{214n}$Ra, $^{216m}$Th, $^{218m}$U in the spin-parity state of 8$^+$, and the shape isomer $^{236m}$U, $^{238m}$U in 0$^+$. They have the excitation energy of about 2 MeV significantly higher than the odd-mass and odd-odd isomers. The isomer $^{212m}$Po, with a half-life of 45.1 s in 18$^+$, is much more stable than the ground state with a half-life of 299 ns. Its corresponding ground state nuclei $^{212}$Po is famous for the $\alpha$ particle moving around the core of $^{208}$Pb. Moreover their preformation probability ratios are all far less than 1, except the shape isomer $^{236m}$U, $^{238m}$U and $^{250m}$Fm without accurate $\alpha$ decay branching ratio, meaning the strong suppression of $\alpha$ clustering in nuclear isomers of even-even nuclei. 

The favored $\alpha$ decay (with l=0), that the nuclear isomer decays to the isomeric state of the daughter nuclei with the same spin-parity, occurs for most odd-mass, odd-odd nuclei isomers. The determination of $\alpha$ decay channel depends on the length of $\alpha$ decay half-life. The shorter decay half-life, the larger branching ratio. The large angular momentum transfer competed with large $\alpha$ decay energy $Q_\alpha$ bring about stability of the nuclei~\cite{Dong10}. For example the $^{209m}$Ra in 13/2$^+$ has the half-time of 117 us and the estimated $\alpha$ decay branching ratio of 90 $\%$ (approximate) measured in 2008~\cite{Aud12}. If it decays to isomeric state $^{205m}$Rn the $Q_\alpha=7.355MeV$ and $l=0$, while if it decays to ground state $^{205}$Rn the $Q_\alpha=8.015MeV$ and $l=5$. The half-life $T_{1/2}$ for the first case equals to 1.15 s and reduces to 0.138 s for the second case. So the second decay channel with $l=5$ dominates the $\alpha$ decay. Considering the coupling of this two channels the total $T_{1/2}=0.123 s$. This is a special example that the relatively high excitation energy of 883 keV provides enough energy for the $\alpha$ particle to get over the additional centrifugal barrier. Many unfavored decays such as $^{209m}$Ra, $^{187}$Bi, $^{254m}$No, $^{214n}$Ra, $^{216m}$Th, $^{218m}$U, $^{212m}$Po, $^{217m}$Pa, $^{214m}$Fr and so on, are similar. It is interesting that the unfavored $\alpha$ decay (with l $\neq$ 0) dominates the process of decaying other than the favored decay. In addition, the numerical results of $\alpha$ decay partial half-life for $^{209m}$Ra are much longer than the experimental data of 130 us, meaning the $\alpha$ decay branching ratio may be estimated too large. The large deviations between the experimental data and calculated results may result from the uncertainty of $\alpha$ decay branching ratio or atomic mass or spin-parity.

\begin{figure}
\centering
\includegraphics[width=9cm]{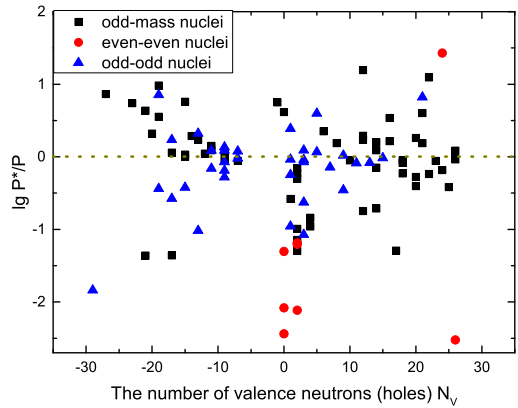}
\caption{(Color online) Logarithmic of the preformation probabilities ratio $P^*_\alpha/P_\alpha$ as a function of the number of valence neutrons (holes)}
\label{Fig1}
\end{figure}

As we all known, the $\alpha$ particle preformation probabilities of even-even nuclei are bigger than those of odd-mass and odd-odd nuclei. If the excitation of single nucleon to a higher energy level occurs, the core of the parent nuclei can be seen as an even-even nuclei. So the preformation probabilities ratios of odd-mass nuclei isomers should be larger than 1. We plot the logarithm of preformation probabilities ratio of isomers to ground states $P^*_\alpha/P_\alpha$ as a function of the number of valence neutrons (holes) N$_V$ in Figure \ref{Fig1}. The black squares and blue triangles represent the ratios of odd-mass nuclei and odd-odd nuclei, respectively. It can be seen that the ratios are greater than 1 for most odd-mass nuclei with valence holes (N$_V$ $<$ 0), which indicates the weak coupling of the excited odd nucleon and the even-even core. And for other odd-mass and odd-odd nuclei, the ratios varies around 1. The red circles indicate the ratios of even-even nuclei, many of which are much less than 1 meaning the strong suppression of $\alpha$ clustering in nuclear isomers. 

Table~\ref{Tab2} is similar to the Table~\ref{Tab1} but for the $\alpha$ transitions where the sequence of energy levels change. With respect to the energy levels, they are closely spaced. So the sequence of energy levels sometimes interchanges. For example in $^{153}$Ho the state of 1/2$^+$ is higher than 11/2$^-$, while in its $\alpha$ decay daughter nuclei $^{149}$Tb the state of 11/2$^-$ is higher than 1/2$^+$. The $\alpha$ particle preformation probabilities ratio of isomers to corresponding ground states are less than 1 for most of these nuclei, meaning the suppression of $\alpha$ particle clustering in nuclear isomers. 

\begingroup
\renewcommand*{\arraystretch}{1.3}
\begin{longtable*}{ccccccccc}
\caption{Same as Table 1, but for the $\alpha$ transitions where the sequence of energy levels change.}
\label{Tab2} \\
\hline 
$\alpha$ transition & $I^{\pi}_i\to I^{\pi}_j$ & $Z_p$ & $N_p$ & $l_{min}$ & $Q_\alpha$ & $T^{exp}_{1/2}(s)$ & $T^{cal}_{1/2}(s)$ & $P^*_\alpha/P_\alpha$ \\ 
\hline
\endfirsthead
\multicolumn{9}{c}%
{{\tablename\ \thetable{} -- continued from previous page}} \\
\hline 
$\alpha$ transition & $I^{\pi}_i\to I^{\pi}_j$ & $Z_p$ & $N_p$ & $l_{min}$ & $Q_\alpha$ & $T^{exp}_{1/2}(s)$ & $T^{cal}_{1/2}(s)$ & $P^*_\alpha/P_\alpha$ \\ 
\hline
\endhead
\hline \multicolumn{9}{r}{{Continued on next page}} \\
\endfoot
\hline \hline
\endlastfoot
$^{151}$Ho$\to^{147m}$Tb&11/2$^{(-)}$$\to$11/2$^{-\#}$ &67&84&0&4.644&$1.60\times10^{2}$&$3.29\times10^{2}$&-\\
$^{151m}$Ho$\to^{147}$Tb&1/2$^{(+)}$$\to$(1/2$^+$)&67&84&0&4.736&$6.13\times10^{1}$&$8.82\times10^{1}$&0.699\\
$^{153}$Ho$\to^{149m}$Tb&11/2$^-$$\to$11/2$^-$&67&86&0&4.016&$2.36\times10^{5}$&$1.84\times10^{6}$&-\\
$^{153m}$Ho$\to^{149}$Tb&1/2$^+$$\to$1/2$^+$&67&86&0&4.121&$3.10\times10^{5}$&$3.48\times10^{5}$&0.144\\
$^{159}$Ta$\to^{155m}$Lu&1/2$^+$$\to$(1/2$^+$)&73&86&0&5.660&$3.06\times10^{0}$&$5.13\times10^{1}$&-\\
$^{159m}$Ta$\to^{155}$Lu&11/2$^-$$\to$(11/2$^-$)&73&86&0&5.744&$1.02\times10^{0}$&$1.87\times10^{0}$&0.11\\
$^{167}$Re$\to^{163m}$Ta&(9/2$^-$)$\to$(9/2$^-$)&75&92&0&5.145&$3.40\times10^{0}$&$1.43\times10^{4}$&-\\
$^{167m}$Re$\to^{163}$Ta&1/2$^+$$\to$1/2$^+$&75&92&0&5.405&$5.90\times10^{2}$&$4.19\times10^{2}$&0\\
$^{170}$Ir$\to^{166m}$Re&(3$^-$)$\to$3$^{-\#}$&77&93&0&5.955&$1.75\times10^{1}$&$3.90\times10^{1}$&-\\
$^{170m}$Ir$\to^{166}$Re&(8$^+$)$\to$(7$^+$)&77&93&2&6.265&$2.25\times10^{0}$&$4.12\times10^{0}$&0.82\\
$^{169}$Re$\to^{165m}$Ta&(9/2$^-$)$\to$(9/2$^-$)&75&94&0&4.989&$1.53\times10^{5}$&$5.23\times10^{4}$&-\\
$^{169m}$Re$\to^{165}$Ta&(1/2$^+$,3/2$^+$)$\to$(1/2$^+$,3/2$^+$)&75&94&0&5.189&$7.55\times10^{3}$&$4.48\times10^{3}$&1.734\\
$^{171}$Ir$\to^{167m}$Re&1/2$^+$$\to$1/2$^+$&77&94&0&5.855&$3.10\times10^{0}$&$4.66\times10^{1}$&-\\
$^{171m}$Ir$\to^{167}$Re&(11/2$^-$)$\to$(9/2$^-$)&77&94&2&6.155&$2.72\times10^{0}$&$7.34\times10^{0}$&0.179\\
$^{173}$Ir$\to^{169m}$Re&(1/2$^+$,3/2$^+$)$\to$(1/2$^+$,3/2$^+$)&77&96&0&5.541&$1.29\times10^{2}$&$8.57\times10^{2}$&-\\
$^{173m}$Ir$\to^{169}$Re&(11/2$^-$)$\to$(9/2$^-$)&77&96&2&5.942&$1.83\times10^{1}$&$2.37\times10^{1}$&0.194\\
$^{186}$Bi$\to^{182m}$Tl&(3$^+$)$\to$2$^{-\#}$&83&103&1&7.655&$1.48\times10^{-2}$&$8.24\times10^{-3}$&-\\
$^{186m}$Bi$\to^{182}$Tl&(10$^-$)$\to$(7$^+$)&83&103&3&7.925&$9.80\times10^{-3}$&$3.39\times10^{-3}$&0.621\\
$^{187}$Bi$\to^{183}$Tl&(9/2$^{-\#}$)$\to$1/2$^{(+)}$&83&104&5&7.779&$3.70\times10^{-2}$&$1.67\times10^{-2}$&-\\
$^{187m}$Bi$\to^{183}$Tl&1/2$^{+\#}$$\to$1/2$^{(+)}$&83&104&0&7.887&$3.70\times10^{-4}$&$5.56\times10^{-4}$&0.174\\
$^{189}$Bi$\to^{185m}$Tl&(9/2$^-$)$\to$9/2$^{-\#}$&83&106&0&6.813&$6.58\times10^{-1}$&$1.39\times10^{0}$&-\\
$^{189m}$Bi$\to^{185}$Tl&(1/2$^+$)$\to$1/2$^{+\#}$&83&106&0&7.452&$9.80\times10^{-3}$&$8.22\times10^{-3}$&0.396\\
$^{191}$At$\to^{187m}$Bi&(1/2$^+$)$\to$1/2$^{+\#}$&85&106&0&7.714&$2.10\times10^{-3}$&$8.05\times10^{-3}$&-\\
$^{191m}$At$\to^{187}$Bi&(7/2$^-$)$\to$9/2$^{-\#}$&85&106&2&7.880&$2.20\times10^{-3}$&$4.44\times10^{-3}$&0.527\\
$^{191}$Bi$\to^{187m}$Tl&(9/2$^-$)$\to$(9/2$^-$)&83&108&0&6.443&$2.43\times10^{1}$&$3.80\times10^{1}$&-\\
$^{191m}$Bi$\to^{187}$Tl&(1/2$^+$)$\to$(1/2$^+$)&83&108&0&7.018&$1.81\times10^{-1}$&$2.33\times10^{-1}$&0.826\\
$^{193}$At$\to^{189m}$Bi&1/2$^{+\#}$$\to$(1/2$^+$)&85&108&0&7.388&$2.90\times10^{-2}$&$1.22\times10^{-1}$&-\\
$^{193m}$At$\to^{189}$Bi&7/2$^{-\#}$$\to$(9/2$^-$)&85&108&2&7.581&$2.10\times10^{-2}$&$5.19\times10^{-2}$&0.589\\
$^{193n}$At$\to^{189n}$Bi&13/2$^{+\#}$$\to$(13/2$^+$)&85&108&0&7.256&$1.17\times10^{-1}$&$2.60\times10^{-1}$&0.903\\
$^{189}$Hg$\to^{185m}$Pt&3/2$^-$$\to$(1/2$^-$)&80&109&2&4.530&$>1.52\times10^{7}$&$1.82\times10^{10}$&-\\
$^{189m}$Hg$\to^{185}$Pt&13/2$^+$$\to$(9/2$^+$)&80&109&2&4.715&$>1.72\times10^{7}$&$1.05\times10^{9}$&0.051\\
$^{193}$Bi$\to^{189m}$Tl&(9/2$^-$)$\to$9/2$^{(-)}$&83&110&0&6.021&$1.82\times10^{3}$&$2.00\times10^{3}$&-\\
$^{193m}$Bi$\to^{189}$Tl&(1/2$^+$)$\to$(1/2$^+$)&83&110&0&6.613&$3.81\times10^{0}$&$6.91\times10^{0}$&1.647\\
$^{195}$Bi$\to^{191m}$Tl&(9/2$^-$)$\to$9/2$^{(-)}$&83&112&0&5.535&$6.10\times10^{5}$&$3.80\times10^{5}$&-\\
$^{195m}$Bi$\to^{191}$Tl&(1/2$^+$)$\to$(1/2$^+$)&83&112&0&6.232&$2.64\times10^{2}$&$2.61\times10^{2}$&1.59\\
$^{197}$Bi$\to^{193}$Tl&(9/2$^-$)$\to$1/2$^{(+\#)}$&83&114&5&5.365&$5.60\times10^{8}$&$6.18\times10^{7}$&-\\
$^{197m}$Bi$\to^{193}$Tl&(1/2$^+$)$\to$1/2$^{(+\#)}$&83&114&0&5.897&$5.50\times10^{2}$&$6.05\times10^{3}$&0.01\\
$^{265}$Sg$\to^{261m}$Rf&9/2$^{+\#}$$\to$9/2$^{+\#}$&106&159&0&8.985&$<1.84\times10^{1}$&$6.95\times10^{0}$&-\\
$^{265m}$Sg$\to^{261}$Rf&3/2$^{+\#}$$\to$3/2$^{+\#}$&106&159&0&9.125&$<2.46\times10^{1}$&$3.57\times10^{0}$&0.384\\
\end{longtable*}
\endgroup

\section{Summary}

In summary, we use the two-potential approach to systematically study the relationship between the nuclear isomers and $\alpha$ particle preformation probabilities. The ratio of $\alpha$ particle preformation probabilities of nuclear isomers to corresponding ground states are extracted from the experimental $\alpha$ decay half-lives. The results indicate the ratios of preformation probabilities vary around 1 for most of isomers, and the strong suppression of $\alpha$ particle clustering for even-even nuclei. We also find for some isomers and ground states the unfavored $\alpha$ decay dominates the process of decaying. 

\begin{acknowledgments}

This work is supported in part by the National Natural Science Foundation of China (Grant No.11205083), the construct program of the key discipline in hunan province, the Research Foundation of Education Bureau of Hunan Province,China (Grant No.15A159 ),the Natural Science Foundation of Hunan Province,China (Grant No.2015JJ3103),the Innovation Group of Nuclear and Particle Physics in USC, Hunan Provincial Innovation Foundation for Postgraduate (Grant No.CX2015B398).

\end{acknowledgments}

\end{document}